\documentclass[english,prd,superscriptaddress,nofootinbib,
               preprintnumbers,twocolumn,noshowpacs]{revtex4}
\usepackage[latin1]{inputenc}
\usepackage{amsmath}
\usepackage{amssymb}
\usepackage{graphicx}
\usepackage[caption=false]{subfig}
\usepackage{amsthm}
\usepackage{bm}
\usepackage{color}
\makeatletter
\@ifundefined{textcolor}{}
{%
 \definecolor{BLACK}{gray}{0}
 \definecolor{WHITE}{gray}{1}
 \definecolor{RED}{rgb}{1,0,0}
 \definecolor{GREEN}{rgb}{0,1,0}
 \definecolor{BLUE}{rgb}{0,0,1}
 \definecolor{CYAN}{cmyk}{1,0,0,0}
 \definecolor{MAGENTA}{cmyk}{0,1,0,0}
 \definecolor{YELLOW}{cmyk}{0,0,1,0}
 }
\usepackage{amsfonts}\usepackage{dcolumn}
\usepackage{babel}

\newcommand{\Mpl}{M_{\rm pl}}

\renewcommand{\O}{\mathcal O}
\newcommand{\sig}{\sigma}
\newcommand{\cosec}{{\rm cosec}}
\renewcommand{\d}{{\rm d}}
\renewcommand{\P}{{\mathcal P}}
\newcommand{\ep}{\epsilon}
\newcommand{\fnl}{f_{\mathrm{NL}}}
\newcommand{\be}{\begin{equation}}
\newcommand{\ee}{\end{equation}}
\def\bs{\begin{subequations}}
\def\es{\end{subequations}}

\begin{document}

\title{What Planck does not tell us about inflation}

\author{Joseph Elliston}
\email{j.elliston@sussex.ac.uk}

\affiliation{Astronomy Centre, University of Sussex,
Falmer, Brighton BN1 9QH, UK.}

\affiliation{School of Physics and Astronomy, Queen Mary University of London,\\
Mile End Road, London, E1 4NS,  UK.}

\author{David Mulryne}
\email{d.mulryne@qmul.ac.uk}

\author{Reza Tavakol}
\email{r.tavakol@qmul.ac.uk}

\affiliation{School of Physics and Astronomy, Queen Mary University of London,\\
Mile End Road, London, E1 4NS,  UK.}

\date{\today}

\begin{abstract}
Planck data has not found the `smoking gun' of non-Gaussianity that would have
necessitated consideration of inflationary models beyond the simplest canonical single
field scenarios. This raises the important question of what these results do imply
for more general models, and in particular, multi-field inflation.
In this paper we revisit four ways in which two-field scenarios can behave differently from
single field models; two-field slow-roll dynamics, curvaton-type behaviour,
inflation ending on an inhomogeneous hypersurface and modulated reheating.
We study the constraints that Planck data puts on these classes of behaviour, 
focusing on the latter two which have been least studied in the
recent literature. We show that these latter classes are almost equivalent, and extend their previous analyses
by accounting for arbitrary evolution of the isocurvature mode which, in particular, places 
important limits on the Gaussian curvature of the reheating hypersurface.
In general, however, we find that Planck bispectrum results only constrain certain regions of parameter space,
leading us to conclude that inflation sourced by more than one scalar field remains an important possibility.

\end{abstract}

\maketitle
        
\section{Introduction} \label{sec:introduction}

Inflation is currently the most promising framework for the dynamics of 
the very early Universe. At present, however, we are some
way from a complete understanding of high energy
physics beyond the Standard Model in which inflation is presumably
embedded. Nevertheless,
attempts have been made to realise
inflation within supersymmetric extensions, and within supergravity
and string theories (for example, see Refs~\cite{Mazumdar:2007vq, Burgess:2013sla}).
No one model is at the present time convincing, but all share a
common feature that the inflationary dynamics is
expected to be sensitive to more than one scalar degree of
freedom~\cite{Quevedo:2002xw,Linde:2005dd,McAllister:2007bg}.
Much of the inflationary literature therefore argues that while
single field models are \emph{simpler}, in the sense that they possess fewer degrees of freedom, 
multiple field models may be more {\it natural}, in the sense that they may be better motivated by fundamental theory.

When more than one light field is present, the dynamics of an inflationary model can 
become extremely rich due to the presence of isocurvature modes. One \emph{possible} consequence of this richness is the generation
of perturbations whose statistics deviate from Gaussian at a much higher
level than the order of slow-roll values that canonical single field models allow \cite{Maldacena:2002vr}.
Planck data, however, has not found evidence for deviations from
Gaussianity \cite{Ade:2013ydc}\footnote{It is important to note that the
current constraints still allow a local bispectrum orders of magnitude larger
than the single field slow-roll value}, and this raises the important question of what this lack of detection
implies for multi-field inflation. 

Here we confine ourselves to two-field models, 
as prototypes of more general multi-field models,
and consider four types of
behaviour which are not possible in single field settings,
and which have been shown to be capable of producing levels
of non-Gaussianity much larger than those possible in canonical
single field models. These consist of models in which
both fields evolve during inflation, leading to a wide spectrum
of possible behaviour~\cite{Bernardeau:2002jy,Rigopoulos:2005us,Vernizzi:2006ve,Alabidi:2006hg,Byrnes:2008wi,Elliston:2011dr, Elliston:2012wm,Tanaka:2010km,Peterson:2010mv,Kim:2010ud,Mulryne:2011ni};
curvaton-type behaviour which we define as the case where a second field begins to oscillate after the first field has 
reheated into radiation and its relative contribution to the energy density can gradually increase\footnote{Our analysis is slightly more general than the standard 
curvaton, or mixed curvaton-inflaton cases which implicitly assume the contribution to the curvaton perturbation from the second field is significant.} \cite{Enqvist:2001zp,Lyth:2001nq, Moroi:2001ct,Enqvist:2005pg,Linde:2005yw,Malik:2006pm,Sasaki:2006kq} (also see Ref.~\cite{Mollerach:1989hu,Linde:1996gt});
and finally scenarios in which the process by which one (or both) of the fields reheat is
directly dependent on the value of one of the light fields, which includes the scenarios known as Modulated
Reheating ({\sc mr})~\cite{Dvali:2003em,Zaldarriaga:2003my,Kofman:2003nx,Vernizzi:2003vs,Bernardeau:2004zz,Ichikawa:2008ne,Byrnes:2008zz,Suyama:2007bg}
and the Inhomogeneous End of Inflation ({\sc iei})~\cite{Bernardeau:2002jf,Lyth:2005qk,Salem:2005nd,Alabidi:2006wa,Bernardeau:2007xi,Sasaki:2008uc,Naruko:2008sq}. 
For reviews of these scenarios see Refs.~\cite{Byrnes:2010em, Suyama:2010uj}.

A particular aim of our work is to show that, 
despite the assertion that Planck data `severely limits the extensions of the simplest paradigm'~\cite{Ade:2013ktc},
multi-field models which exhibit the behaviour described above only produce an inconsistent level of 
non-Gaussianity for (often very) restricted ranges of their possible 
initial condition and parameter ranges. As we will see, this conclusion is obvious 
for the curvaton-type case and for models where both fields evolve during inflation, as these generically 
rely on `finely tuned' initial conditions in order to produce a level of non-Gaussianity inconsistent with data. 
Moreover, these cases have been studied in detail in recent years \cite{Byrnes:2010em,Elliston:2011dr} and 
in particular since the Planck results~\cite{Enqvist:2013paa, Kobayashi:2013bna}.
To gain further understanding of the consequences of the Planck constraints for
two-field models, we therefore focus on the {\sc iei} and {\sc mr} scenarios.
In order to form a more complete understanding of the restrictions Planck data puts on these models, 
we study them in a geometric way that allows us
to present the first calculation that includes evolution of the 
isocurvature mode during inflation. 
This extends previous geometric approaches by Naruko and Sasaki~\cite{Naruko:2008sq}, Huang~\cite{Huang:2009vk}
and Matsuda~\cite{Matsuda:2012kk}.
Using this broader framework we show, in a precise way, that
the {\sc mr} and {\sc iei} scenarios are almost, but not exactly, identical. 
This broader framework also highlights the physical processes that allow a large {\sc cmb} bispectrum
signal to occur. 
Our study is complementary to other works which have highlighted the relations between various two-field models \cite{Vernizzi:2003vs, Alabidi:2010ba}.

The outline of the paper is as follows: \S\ref{sec:theory} covers background material and in 
\S\ref{sec:FSM} we give a brief discussion of the four scenarios discussed above.
In \S\ref{sec:general} we discuss the evolution of $\zeta$ and isocurvature in the {\sc mr} and {\sc iei} scenarios.
\S\ref{sec:equivalence} then presents the analytic predictions for the {\sc mr} and {\sc iei} scenarios, 
demonstrating the degree to which they are equivalent as well as discussing
the conditions required for them to generate a large {\sc cmb} bispectrum. 
To concretize the level of tuning required for the production of non-Gaussianity in these scenarios we then consider
two representative models in \S\ref{sec:models},
where we study their parameter spaces by simultaneously imposing the Planck bounds
on the spectral index and the bispectrum. Finally we conclude in \S\ref{sec:conclusions}.

\section{Background theory} 
\label{sec:theory}

The most important observational signatures of an inflationary model are the statistics of the scalar curvature perturbation 
on uniform density hypersurfaces, denoted $\zeta$, and of the tensor perturbations also produced by inflation. 
The two-point function of $\zeta$ is parametrized by the power spectrum $\P_\zeta (k)$, the weak scale  
dependence of which is given by the spectral index, 
$n_\zeta -1 \equiv \d \ln \P_\zeta (k) / \d \ln k$.
About some pivot scale $k^*$, taken to be close to the largest observable scale,
the amplitude of the scalar power spectrum and the spectral index have been observationally determined to high accuracy. 
Current data gives $\P_\zeta = {2.196}^{+0.051}_{-0.060} \times 10^{-9}$,
 and $n_\zeta = 0.9616\pm 0.0094$ at $68$\%~{\sc cl} \cite{Ade:2013zuv}.
Tensor perturbations remain undetected, with the current constraint 
on the ratio of the tensor to the scalar power spectra at the pivot scale given by $r < 0.11$ at 95\% {\sc cl}~\cite{Ade:2013zuv}. 
Likewise no evidence for the primordial distribution of $\zeta$ being  
non-Gaussian has yet been detected. The first non-Gaussian statistic 
which could potentially be observed is the three-point 
function of $\zeta$, the prediction of which is often written in terms of the reduced bispectrum $\fnl(k_1, k_2, k_3)$. 
For the models considered in this paper, only the local shape bispectrum is generated at a level which could possibly 
be observed and this is simply a number. For brevity we denote it $\fnl$.  Planck constraints are  
$\fnl = 2.7 \pm 5.8$ at $68$\%~{\sc cl}~\cite{Ade:2013ydc}. \\

\noindent \textbf{\textit{Separate universe picture.}} 
The separate universe picture asserts that when considering the perturbed Universe smoothed on scales much greater than the 
horizon size, spatial gradients may be neglected, and smoothed patches evolve as
independent `separate universes' \cite{Lyth:1984gv,  Starobinsky:1986fxa, Wands:2000dp}. 
Local quantities, such as pressure, density, and the amount of expansion can be 
different in different patches, but each patch evolves according to the background equations of motion
for a Friedman-Lema\^{i}tre-Robertson-Walker ({\sc flrw}) universe.\\

\noindent\textbf{\textit{Phase space bundle.}} 
For inflationary applications, the initial conditions for each separate universe are usually fixed on a flat slicing of spacetime at 
the time of horizon crossing, and are perturbatively different for each universe.
Statistical homogeneity of our Universe implies that we don't care about the spatial position of individual separate universes, 
but only about the statistics of local quantities, such as the 
variance of the expansion undergone by the universes. 
These properties can be related to the statistics of $\zeta$ through the $\delta N$ formalism \cite{Sasaki:1995aw, Lyth:2005fi}, as we describe below. 

Given that we only wish to track information about the local properties in each separate universe, and not 
spatial information, the universes can be mapped onto trajectories in a phase space  
defined by the {\sc flrw} system.
The ensemble of separate universes can then be intuitively described as a `bundle' 
in phase space (for a more detailed description see for example Refs.~\cite{Elliston:2011et, Seery:2012vj} and references therein),  and 
the evolution of this bundle encodes all the information we require about the perturbed Universe.

During slow-roll inflation, which applies around horizon exit, the evolution of a scalar 
field approaches an attractor solution
\be
3 H \dot \phi_i \simeq V_{,i} \,,
\label{eq:scaling}
\ee
with $\dot \phi_i = \d \phi_i / \d t$ and $3 \Mpl^2 H^2 = \rho$, where $\rho$ and $V$ are the density and the inflationary potential respectively and $\rho\approx V$ during slow roll.
Under this attractor condition it is clear that $\phi_i$ and $\dot \phi_i$ are not independent and so the effective 
dimensionality of the phase space is reduced.

It then proves convenient to define the potential derivatives $V_{,i} = \partial V / \partial \phi_i$ in terms of the potential slow-roll parameters
\be
\label{eq:sr_parameters}
\epsilon_i = \frac{\Mpl^2}{2} \frac{V_{,i}^2}{V^2}\,, \quad 
\eta_{ij} = \Mpl^2 \frac{V_{,ij}}{V}\,, \quad
\xi_{ijk}^2 = \Mpl^3 \sqrt{2 \ep} \frac{V_{,ijk}}{V}\,,
\ee
where $\ep = \sum_i \ep_i$.
For the application of eq.~\eqref{eq:scaling} we may pick, without loss of generality, $\dot \phi_i <0$. 
This then implies that we take the positive branch when taking the square root of $\ep_i$ such that $\Mpl\, V_{,i} = +V \sqrt{2 \ep_i}$.\\

\noindent\textit{\textbf{$\bm{\delta N}$ formalism.}} 
It is helpful to evaluate the separate universes described above at some time after horizon crossing 
at which they all share the same energy density, which defines a uniform density slicing of spacetime. 
In phase space, this defines a uniform density hypersurface, the geometry of which is inextricably linked to the inflationary dynamics, 
since it lies perpendicular to the direction of evolution of the bundle.
Each separate universe, when compared to some fiducial member, will reach this uniform density hypersurface at a slightly different time $\delta N$.
The `$\delta N$ formalism' then prescribes that $\zeta = \delta N$.
In the presence of multiple light fields one expands $\delta N$ perturbatively via a Taylor expansion 
in the horizon crossing field values as
\be
\label{eq:deltaN}
\zeta = \delta N = N_{,i} \delta \phi^*_i +  \frac{1}{2} N_{,ij} \delta \phi^*_i\delta \phi^*_j + \dots ,
\ee
where `$*$' indicates evaluation on the flat hypersurface at horizon crossing and $N_{,i} = \partial N / \partial \phi_i^*$. 
We will utilise the label `$c$' for quantities evaluated on the final uniform density hypersurface. \\

\noindent\textit{\textbf{The evolution of $\bm \zeta$.}}
Since $\zeta$ is additive, we may simplify some of the discussion in this paper
by decomposing it into three components as $\zeta = \zeta_{\rm hor} + \zeta_{\rm inf} + \zeta_{\rm reh}$,
where $\zeta_{\rm hor}$ is the contribution present at horizon crossing, $\zeta_{\rm inf}$ is the contribution
arising during inflation and $\zeta_{\rm reh}$ is the contribution arising from the mechanism by which the Universe reheats. 
For a system with two fields $\phi$ and $\sig$, this decomposition of $\zeta$ is particularly useful if we choose a field basis such that $\phi$ 
is the adiabatic field at horizon exit (i.e. it is aligned with the direction of phase space flow), and $\sig$ is the perpendicular isocurvature field.

On the flat hypersurface near horizon exit, both fields have Gaussian field perturbations \cite{Seery:2005gb} $\{\delta \phi^*,\delta \sig^*\}$.
One then finds $\zeta_{\rm hor}$ by computing the excess expansion $\delta N$ that puts the separate universes onto a nearby
uniform density hypersurface where $\delta \phi = 0$.  This means that, to second order, $\zeta_{\rm hor}$ obeys the single field result 
\be
\label{eq:zeta_hor}
\zeta_{\rm hor} = \frac{1}{\sqrt{2 \ep^*}} \frac{\delta \phi_*}{\Mpl}
+\frac{1}{2} \bigg( 1- \frac{\eta_{\phi \phi}^*}{2 \ep^*} \bigg) \frac{\delta \phi_*^2}{\Mpl^2} + \O(\delta \phi_*^3)\,.
\ee
From this expression we may read off the `$\delta N$ coefficients' $N_{,\phi}$ and $N_{,\phi \phi}$ and these remain constant regardless of the subsequent evolution.
Conversely, the perturbations $\delta \sig^*$ are unaltered by this gauge transformation and so $N_{,\sig}$, $N_{,\sig \sig}$ and $N_{,\phi \sig}$ are all zero at horizon exit,
although one expects these to evolve subsequently. If these evolve during inflation then this will source a contribution to 
$\zeta_{\rm inf}$ and/or if these evolve at the time of reheating then they will contribute to $\zeta_{\rm reh}$.\\

\noindent\textit{\textbf{Cosmological parameters.}}
We may use the $\delta N$ formalism to write simple expressions for the cosmological 
parameters $\P_\zeta$, $n_\zeta$, $r$ and $\fnl$. It proves convenient to write these in terms of the parameter $R=N_{,\sig}^2/N_{,\phi}^2$
such that, at linear order, $R \ll 1$ implies that the $\phi$ field effects dominate and $\zeta \approx \zeta_{\rm hor}$,
whereas $R \gg 1$ implies that the $\sig$ field effects dominate and $\zeta \approx \zeta_{\rm inf} + \zeta_{\rm reh}$.
Both fields contribute equally to $\zeta$ when $R=1$. 
This same notation was recently applied to the curvaton scenario in Ref.~\cite{Enqvist:2013paa}.
We therefore see that $R=0$ at horizon crossing, after which $R$ will evolve; this evolution may involve both growth and decay of $R$,
but $R$ always remains positive definite.
Restricting to sum-separable potentials which do not possess direct couplings between
the two fields, one finds~\cite{Sasaki:1995aw,Vernizzi:2006ve}
\bs
\begin{align}
\P_\zeta &= \frac{1}{2 \ep^*} \big(1 + R\big) \Mpl^{-2} \P_{\delta \phi} \,, 
\label{eq:power} \\
r &= \frac{16 \ep^*}{1+R} \,, \label{eq:r} \\
n_\zeta -1 &=
\frac{2R (\eta_{\sig \sig}^* - \ep^*)  + 2\eta_{\phi \phi}^* - 6 \ep^*}{1+R} \,,
\label{eq:ns} \\
\frac{6}{5} \fnl &\supseteq \frac{R^2}{(1+R)^2} 
\frac{N_{,\sig \sig}}{N_{,\sig}^2}
\label{eq:fnl} \,,
\end{align}
\es
where $\P_{\delta \phi} = \langle \delta \phi^* \,\delta \phi^* \rangle = 
\langle \delta \sig^* \,\delta \sig^* \rangle = H_*^2 / 4\pi^2$.
Eq.~\eqref{eq:fnl} is only valid if $\fnl$ is `large'
(i.e. the observationally relevant regime where $|\fnl|\geq \O(1)$), and in other settings one finds $\fnl = \O(\ep^*)$. \\

\noindent\textit{\textbf{Constraint on bundle width.}}
The power spectrum $\P_\zeta$ is the only cosmological parameter that is proportional to the energy scale of inflation through $H^*$.
As a result, the value of $\P_\zeta$ can be fixed by picking the normalisation of the inflationary potential.
However, we point out that there is another subtle constraint that can be derived from the power 
spectrum by noting that it informs us about the width of the bundle on a uniform density hypersurface `$c$' at some later time during its evolution.
To show this, it proves convenient to define a new basis that is aligned with the adiabatic and isocurvature fields 
at the later time in question, with $\tilde \sig^c$ as the local isocurvature field. The bundle width at this time is therefore denoted
$\delta \tilde \sig^c$ and may be calculated as
\be
\delta \tilde \sig^c = \frac{\partial \tilde \sig^c}{\partial \sig^*} \sqrt{\frac{2 \ep^* \Mpl^2 \P_\zeta }{1+R}}.
\ee
When we look at particular models in \S\ref{sec:models} we shall impose two constraints on this bundle width. 
The first constraint is that the absolute value of $\delta \tilde \sig$
does not ever grow sufficiently large that the bundle picture becomes invalid. To this end we require that $\delta \tilde \sig < 0.1 \, \Mpl$.
The second constraint is that the bundle width is much smaller than the characteristic scale describing any feature
associated with the inflationary potential, such as the radius of curvature of the reheating hypersurface. 
This assumption is made to eliminate the more complex possibility that the bundle may bifurcate, which then requires the different branches of this bifurcation 
to follow almost identical expansion histories to avoid violating the observed statistical homogeneity.
This possibility has been considered elsewhere~\cite{Li:2009sp,Wang:2010rs,Duplessis:2012nb}. 
For the models that we consider in \S\ref{sec:models}, we find that this second constraint is always satisfied.

\section{Four classes of models commonly considered}
\label{sec:FSM}

Our four classes of models sample the spectrum of possible behaviours for the evolution of $\zeta$
for inflationary models with two scalar degrees of freedom.
In discussing `two-field slow-roll inflation' we consider two light fields that {\it both} evolve during inflation, causing $\zeta$ to evolve during inflation too.
The remaining three classes of models do not alter $\zeta$ during inflation, but rather cause $\zeta$ to evolve {\it after} inflation has ended,
or through the process by which inflation ends. For simplicity we shall consider these effects independently, 
but one should be mindful that combinations are a possibility (for examples, see Refs.~\cite{Choi:2012te,Assadullahi:2013ey,Langlois:2013dh,Assadullahi:2012yi}).

The three classes of models for which $\zeta$ evolves after inflation are distinguished by the parameters that define their effective phase spaces,
be they fields or fluids (fluids may include radiation, or a scalar field that is oscillating about a potential minimum such that its 
time-averaged behaviour may be described by a fluid). 
The case of curvaton-type dynamics has an effective phase space described by two fluids, 
the {\sc mr} scenario is described by one fluid and one (non-oscillating) scalar field and 
the {\sc iei} scneario is described by two (non-oscillating) scalar fields.
We give full details of these cases in the respective sections below.

For the first two classes, `two-field slow-roll' and the `curvaton-type', we summarise some well established results in simple 
cases, and discuss the implications of Planck data for them. For the second 
two classes, the {\sc mr} and {\sc iei} scenarios, this section serves as a brief introduction to the 
the rest of the paper, which concentrates on 
extending earlier analysis and confronting them with Planck constraints.

\subsection{Two-field slow-roll inflation}
\label{sec:twofield}

This section summarises how $\zeta$ may be generated during inflation sourced by the joint slow-roll evolution of two scalar fields.
As discussed at length in Refs.~\cite{Elliston:2011dr,Elliston:2011et}, $\zeta$ may be generated transiently as the bundle responds to particular
features in the inflationary potential such as ridges, valleys and inflection points. Subsequently, if the potential
allows for quenching of isocurvature perturbations, then $\zeta$ will become conserved~\cite{Rigopoulos:2003ak,Lyth:2004gb} as one approaches an adiabatic limit.
A large non-Gaussianity is possible either during the transient evolution or in an adiabatic limit, but both scenarios require a high degree of fine tuning.
This is easily visualized through the heat map approach of Byrnes {\it et al.} \cite{Byrnes:2008wi},
later adopted and simplified in Ref.~\cite{Elliston:2012wm}. This
approach relies on an analytic expression for $\fnl$ that is only available under the assumptions of 
slow-roll and a separable potential. However, numerical studies~\cite{Mulryne:2011ni,Leung:2012ve} of non-separable potentials
appear to agree with the conclusion that a large $\fnl$ requires very specific initial conditions. Additionally, work on the 
transport approach to inflationary perturbations gives an alternative perspective on the conditions required for a large non-Gaussianity 
by considering the statistics of field perturbations on a flat hypersurface at the end of inflation, and the transformation which converts these to the 
statistics of $\zeta$ \cite{Mulryne:2013uka,Anderson:2012em,Seery:2012vj,Mulryne:2010rp,Mulryne:2009kh}.

\begin{figure}[h]
\centering
\includegraphics[width=\columnwidth]{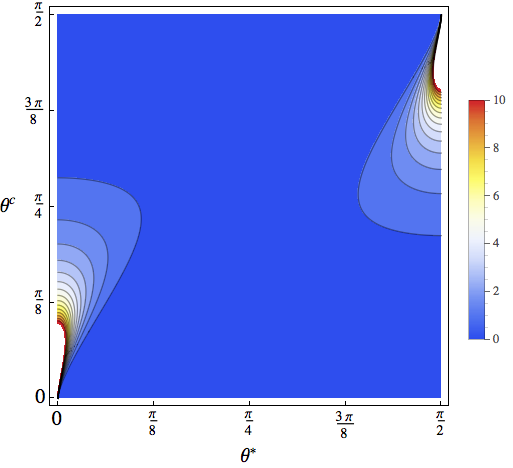}
\caption{Heatmap of the function $f$ on a linear scale. Only the `hot' 
regions can produce a detectable $\fnl$, and the white region has saturated the scale. 
Models begin on the diagonal $\theta^c = \theta^*$ and then evolve vertically only.}
\label{fig:f}
\end{figure}

If the inflationary potential is product-separable or sum-separable with a dominant 
constant term (such as the valley potential example considered below), then $\fnl$ can be written as~\cite{Elliston:2012wm}
\be
\frac{6}{5} \fnl \simeq f\, \Big[2\eta_{\tilde \sig \tilde \sig}^c - \eta_{\sig \sig}^* \Big]  \,,
\label{eq:fnl_simple}
\ee
where the isocurvature $\eta$-parameters are computed at the different times 
`$c$' and `$*$' corresponding to isocurvature fields $\tilde \sig$ and $\sig$ respectively.
This result holds to an excellent degree of approximation in the limit where $\fnl$ is large enough to detect.
An analogous result holds for general sum-separable potentials~\cite{Elliston:2012wm}, but for brevity we shall not 
discuss this case here. 
Since $\eta_{\sig \sig}$ is a slow-roll parameter, a necessary (though not
sufficient) condition for a large $\fnl$ is that $f \gg 1$. $f$ depends on
the angle of phase space velocity, $\theta$, defined with respect to the field basis in which the potential is separable as
\be
\label{eq:f}
f = \frac{\sin^2 2\theta^c (\cos^2 \theta^c - \cos^2 \theta^*)^2}
{4 (\cos^4 \theta^c \sin^2 \theta^* + \sin^4 \theta^c \cos^2 \theta^*)^2} \,.
\ee
We plot $f(\theta^c,\theta^*)$ in 
Fig.~\ref{fig:f}. Since $\theta^*$ is a constant for any inflationary model, 
one immediately sees a large non-Gaussianity can only arise if the initial 
conditions are such that the initial motion is highly aligned to 
one of the principal axes where $\theta^* = 0,\pi/2$. If the 
hierarchy between the masses of the two evolving fields is not large (which is required for both fields to evolve without slow-roll
being violated) then this represents a severe fine tuning of initial conditions~\cite{Byrnes:2008wi}. 
(Note that the two limits $\theta^* = 0,\pi/2$ are equivalent up to an arbitrary field 
relabelling, so one may consider the case with $\theta^* \ll 1$ without loss of generality). 
Physically, this fine tuning is always associated with requiring the phase space bundle to originate in close proximity to
special features in the potential, such as ridges, uplifted valleys or inflection points~\cite{Elliston:2011dr, Elliston:2011et, Elliston:2012wm}, which leads to 
non-linear distortion of the phase space bundle.
In addition, further tuning is required to ensure that inflation ends at the right time to ensure that $\theta^c$ 
accesses the `hot' region at the end of the evolution. 

As a simple example, let us consider the potential
$V \simeq V_0 + \frac{1}{2} m_\phi^2 \phi^2 + \frac{1}{2} m_\sig^2 \sig^2$, 
where we presume that the vacuum term $V_0$ dominates throughout. This 
can be thought of as a Taylor expansion of a more general model.
A simple way of attaining $\fnl \gg 1$ is to have similar but not equal masses with $m_\sig < m_\phi$ and an 
initial condition $\sig^* \ll \Mpl$ with $\phi^* \sim \Mpl$. 
These initial conditions give $\theta^* \ll 1$ such that $\sig$ is approximately the isocurvature field.
However, from Fig.~\ref{fig:f} we see that $\theta$ must grow a small amount
in order for $|\fnl| \gg 1$ to be possible, requiring that the bundle begins to turn. 
As discussed in Ref.~\cite{Elliston:2011dr}, $\fnl$ grows to a large positive value in the early stages of such a turn, but if we allow 
the turn to become even half-way complete then $\fnl$ decays to a value $\sim \O(\ep)$.
We find the peak value of $\fnl$ to occur when $\theta \ll 1$ and $R=3$, meaning that both $\phi$ and $\sig$ contribute roughly equally to $\zeta$,
and we find
\be
\label{eq:peak_fnl}
\left.\frac{6}{5} \fnl \right|_{\rm peak} = \frac{3 \sqrt{3}}{16}\frac{\eta^*_{\sigma \sigma}}{\theta*} \,.
\ee
Not only does a large peak value require fine tuning of $\theta^* \ll 1$, an additional fine tuning is required to ensure that 
inflation ends in the region of the peak. 

Through this example, and the general heatmap analysis, we therefore conclude that the Planck bispectrum bounds 
only provide a very small constraint on the parameter space of two-field slow-roll inflation.
We note that the toy example that we have considered produces a blue tilted spectrum. In more complex cases, however, with a similar level of fine-tuning, 
it is possible construct models of slow-roll inflation which have a reasonable spectral index and a large non-Gaussianity.

\subsection{Curvaton-type behaviour}
\label{sec:curvaton}

Curvaton-type behaviour involves an inflaton field $\phi$ that drives inflation before reheating
into radiation of density $\rho_\phi$. The `curvaton' is a spectator field $\sig$ during inflation with negligible energy density and, 
if we presume a simple quadratic potential $V(\sig)$,
negligible contribution to $\zeta$ before it begins oscillating about the minimum when $\sig = \sig_{\rm osc}$.
During these oscillations in the quadratic minimum, its averaged energy density decreases like a pressureless fluid $\rho_\sig$, 
and so its energy redshifts more slowly than that of the radiation. It therefore 
slowly starts to dominate the dynamics of the universe, and 
$\zeta_{\rm reh}$ grows. At some point, the curvaton also decays into radiation and $\zeta$ becomes conserved.
If $\zeta_{\rm reh}$ grows to dominate $\zeta_{\rm hor}$ at linear order, then this defines the standard curvaton scenario,
whereas in this section we consider the more general case where the inflaton perturbations may be non-negligible or even completely dominant. This has been 
referred to as the mixed curvaton-inflation scenario and studied in Refs~\cite{Langlois:2004nn,Fonseca:2012cj}.  But even that description implies that perturbations 
in the curvaton contribute significantly to $\zeta$. This need not be the case even if the `curvaton' field comes to completely dominate the 
energy density which is why we adopt the curvaton-type label.
This scenario has more recently been considered in detail in Refs.~\cite{Kobayashi:2013bna,Enqvist:2013paa}.

{\it Under what conditions can $\zeta_{\rm reh}$ dominate $\zeta$?}
At linear order one finds~\cite{Lyth:2005fi} $\zeta_{\rm reh} =\frac{2}{3} \frac{r H^*}{\sigma_{\rm osc}} \frac{\partial \sigma_{\rm osc}}{\partial \sigma_*}$, 
where $r = 3 \rho_\sig / (3 \rho_\sig + 4 \rho_\phi)$ is bounded between zero and unity. 
For this to dominate $\zeta_{\rm hor}$ requires quite specific initial conditions as first pointed out in Ref.~\cite{Bartolo:2002vf}, that mirror the fine tuning 
seen in the previous section on two-field slow-roll inflation. For example, if both fields evolve in quadratic potentials,
$\zeta_{\rm reh} \gtrsim \zeta_{\rm hor}$ requires  $\sigma^* \lesssim 0.1 \, \Mpl$ (where we have further presumed $\sig_{\rm osc} \sim \sig^*$),
whereas we require $\phi^* \simeq 16 \Mpl$. This disparity comes into sharper focus when we recall that the inflaton must be must heavier at horizon exit
and hence will evolve to its minimum more rapidly than the curvaton field.
This disparity in initial conditions demonstrates a measure of fine tuning in this model.

Requiring that $\fnl$ is sufficiently large to be constrained by Planck bispectrum data
exacerbates this fine tuning. The evolution of $\fnl$ in this standard quadratic curvaton model closely mirrors how $\fnl$ 
evolves in the two-field slow-roll example considered in \S\ref{sec:twofield} with an uplifted valley potential. Specifically, we again find the peak in
$\fnl$ is positive, it occurs when $R=3$ and the peak value of $\fnl$ is identical to that given in Eq.~\eqref{eq:peak_fnl}.
If we pick $\eta^*_{\sigma \sigma} \sim 0.01$, this model is only
capable of producing a large bispectrum peak for $\sigma^* \lesssim 0.01 \, \Mpl$.\footnote{
The main difference between these two models is the limiting value of $\fnl$ in the limit $R \to \infty$,
if indeed the model is capable of producing large values of $R$. 
In this limit one finds $\fnl \sim \O(\ep)$ for the two-field slow-roll case and $\fnl = -5/4$ for the standard curvaton.}
This tuning is increased if we demand a smaller value of $\eta^*_{\sigma \sigma}$. 
Furthermore, one also needs to ensure that the curvaton reheats in the vicinity of this peak in $\fnl$,
which requires additional tuning of the model parameters.

We conclude that curvaton-type dynamics remain an interesting possibility for 
two-field models. It is clear, however, that for such models to produce a level of 
$\fnl$ which is constrained by Planck requires a similar level of fine tuning 
to that required by the pure inflationary models discussed in the last 
subsection. Of course, in both cases, some dynamical process before the observable inflationary 
phase might set these initial conditions in a natural way, but even this cannot
remove the tuning that enables reheating to occur when $\fnl$ is near its peak value. We therefore
find that such models are not challenged by Planck bispectrum data, except for rather specific choices of initial conditions.

\subsection{Inhomogeneous end of inflation models}
\label{sec:endofinflation}

The {\sc iei} scenario~\cite{Bernardeau:2002jf,Lyth:2005qk,Alabidi:2006wa} can be thought of as a straightforward 
generalisation of the $\delta N$ picture we have described in \S\ref{sec:theory}, where the bundle 
evolves from a flat initial hypersurface to a final uniform density one. 
Instead, the {\sc iei} scenario ends inflation on a hypersurface of arbitrary geometry, and $\zeta_{\rm reh}$ is generated by
the additional $\delta N$ taken for different members of the bundle to reach this end of inflation hypersurface. In principle there will be a contribution from 
after the transition too, as we must ultimately evaluate $\zeta$ on a constant density hypersurface, but this contribution 
is sub-dominant.  
Physically, such a sudden end to inflation can be achieved, for example, at a hybrid-like transition.  
Trajectories before the transition
undergo slow-roll evolution, but inflation ends abruptly at the transition.
The simplest realisation involves an inflaton $\phi$ that determines both the Hubble rate $H(\phi)$ and the number of efoldings $N(\phi)$~\cite{Lyth:2005qk},
and an end of inflation hypersurface, the geometry of which is prescribed by an independent isocurvature field $\sig$.

As noted by Huang~\cite{Huang:2009vk}, one possible source of non-Gaussianity in this case can be understood in terms of the geometric
properties of the end hypersurface, with $\fnl$ being proportional to its Gaussian curvature. In the next section, we shall generalise this scenario,
allowing the perturbations $\delta \sig$ to evolve during inflation which leads to important new effects, including
a secondary mechanism for generating large $\fnl$ that is related to the non-linear evolution of the $\sig$ 
field perturbations.

\subsection{Modulated reheating}
\label{sec:modulated}

The {\sc mr} scenario~\cite{Dvali:2003em,Zaldarriaga:2003my,Kofman:2003nx,Bernardeau:2004zz} shares a number of similarities with the {\sc iei} case.
In its simplest realisation, a scalar field $\phi$ is employed to generate inflation, which ends gracefully as the inflaton reaches the quadratic minimum of its potential.
Subsequently, the $\phi$ field oscillates in this quadratic minimum such that it behaves as a pressure-less fluid. 
Reheating then occurs, and in the simplest picture of perturbative reheating one defines a constant decay width $\Gamma$
such that the fluid $\rho_\phi$ decays quickly into radiation when $H = \Gamma$.
The {\sc mr} scenario is a simple generalisation, promoting the constant parameter $\Gamma$ to become a function $\Gamma(\sig)$ where
$\sig$ is some light scalar field~\cite{Bassett:2005xm,Frolov:2010sz}. The time of reheating for any given separate universe is now dependent on the local value of $\sig$.
Since the energy density decays faster after reheating, this leads to a variation in the number of efolds taken to reach a future uniform density hypersurface
and so an additional contribution to $\zeta_{\rm reh}$ arises.

\section{mr and iei scenarios: evolution of $\bm \zeta$ and isocurvature} 
\label{sec:general}

Our calculation for $\zeta$ in the {\sc iei} and {\sc mr} scenarios is split into three additive parts, corresponding to the contributions
$\zeta_{\rm hor}$, $\zeta_{\rm inf}$ and $\zeta_{\rm reh}$. The first of these terms is always present and was computed in Eq.~\eqref{eq:zeta_hor}.
Since these three contributions to $\zeta$ decouple, and since our interest is in $\zeta_{\rm reh}$ rather than $\zeta_{\rm inf}$,
we are motivated to consider the simplified limit where the phase space bundle follows a straight line such that $\zeta_{\rm inf}=0$. 
If one wishes to consider more general scenarios where the bundle turns during inflation, then $\zeta_{\rm inf}$ may be calculated either
numerically or, in certain scenarios, analytically~\cite{Vernizzi:2006ve,Choi:2012hea}.

This is not, however, the complete story. Whilst the observable predictions (i.e. the contribution to $\zeta$) from each of these three regimes are all additive,
the underlying physics of each regime is {\it not} decoupled. This is due to the presence of an isocurvature perturbation 
which is sensitive to the inflationary dynamics and has the capacity to influence the process of reheating, and thus alter $\zeta_{\rm reh}$.
Following the notation of \S\ref{sec:theory}, where $\delta \tilde \sig^c$ is the isocurvature perturbation on a uniform density 
hypersurface taken just before reheating commences, and $\delta \sig^*$ 
is the isocurvature perturbation near horizon exit, the fact that isocurvature can evolve during inflation means that
$\delta \tilde \sig^c$ is a non-trivial function of $\delta \sig^*$. To account for such behaviour up to second order, as required
for the computation of the bispectrum, we may expand $\delta \tilde \sig^c$ as
\be
\label{eq:isocurvature_exp}
\delta \tilde \sig^c = \frac{\partial \tilde \sig^c}{\partial \sig^*} \delta \sig^* + \frac{1}{2} \frac{\partial^2 \tilde \sig^c}{\partial \sig_*^2} \delta \sig_*^2 + \O(\delta \sig_*^3).
\ee
Note that we do not need to include derivatives with respect to $\phi^*$ because this is defined as the adiabatic field at horizon exit
and so perturbations $\delta \phi^*$ have no effect after horizon exit.
We therefore see that, up to second order, there are two physical degrees of freedom associated with the evolution of isocurvature:
Firstly, the term $\partial \tilde \sig^c / \partial \sig^*$ represents the growth or decay of the width of the bundle, and secondly, the term
$\partial^2 \tilde \sig^c / \partial \sig_*^2$ informs us about the generation of any non-linear growth of isocurvature.

Consequently, when we compute $\zeta_{\rm reh}$ in \S\ref{sec:equivalence}, we can be sure that we are accounting for
an arbitrary evolution of isocurvature by simply incorporating the two terms appearing in the expansion~\eqref{eq:isocurvature_exp}.
Importantly, we note that the inclusion of these two terms is not contingent on the value of $\zeta_{\rm inf}$ and so we can 
make the simplifying assumption that $\zeta_{\rm inf}=0$ without running the risk of
inadvertently ignoring pertinent physical degrees of freedom in $\zeta_{\rm reh}$.
Furthermore, the assumption that the bundle evolves in a straight line means that we are able to easily compute the derivatives such as $\partial \sig^c / \partial \sig^*$,
where the assumption of straight-line evolution means that the tilde on $\sig^c$ is now redundant and has therefore been dropped.\\

\noindent \textbf{\textit{Isocurvature evolution for a straight bundle.}} 
Evolution equations for field perturbations such as $\delta \sig$ can be formulated by perturbing the equations 
of motion for the background evolution, once they are written with the correct time variable \cite{Yokoyama:2007uu, Mulryne:2010rp}. In the present case we wish to follow the evolution 
$\delta \sig^* \to \delta \sig^c$, and this can be done by perturbing Eq.~\eqref{eq:scaling} with respect to the $\sig$ field 
($H$ and $N$ need not be perturbed since the assumption that the bundle evolves in a straight line requires these to be functions of $\phi$ only) and then integrating. One finds
\bs
\begin{align}
\label{eq:isocurvature1}
\frac{\partial \sig^c}{\partial \sig^*} &= \mbox{exp}\left[ -\int_*^c \frac{V_{,\sig \sig}}{3 H^2} \, \d N \right] \,, \\
\label{eq:nonlinear1}
\frac{\partial^2 \sig^c}{\partial \sig_*^2} &= -\frac{\partial \sig^c}{\partial \sig^*} \int_*^c \frac{V_{,\sig \sig \sig}}{3 H^2} \frac{\partial \sig}{\partial \sig^*}
\, \d N \,.
\end{align}
\es
These results tell us that the derivative $\partial \sig^c / \partial \sig^*$ only varies from unity if $\eta_{\sig \sig} \neq 0$ and 
$\partial^2 \sig^c / \partial \sig_*^2$ is only non-zero if $\xi_{\sig \sig \sig}^2 \neq 0$.
The latter condition requires that the potential $V(\sig)$ is not an even function, 
and so we know that the non-linear effect will be absent if the $\sig$ potential is, 
for example, a quadratic, quartic or axion (cosine) form. 
The simplest potential that will generate non-linear isocurvature evolution is an 
inflection point, $V(\sig) = \lambda \sig^3 / 6$.

\section{The Modulated Reheating and Inhomogeneous End of Inflation Equivalence}
\label{sec:equivalence}

We now present the analytic predictions for the {\sc iei} and {\sc mr} scenarios which 
will allow us to show precisely in what ways these scenarios are equivalent and in which ways they differ.
The aim of this section is to understand the physical processes at play in these systems,
and in particular to identify how a large $\fnl$ may develop.
Subsequently, in \S\ref{sec:models}, we shall embed the {\sc iei} and {\sc mr} behaviour in two representative inflationary scenarios 
and constrain these with current data.

The only factors in eqs.~\eqref{eq:power} to \eqref{eq:fnl} that relate to the end of inflation effects are $R$ and $N_{,\sig \sig} / N_{,\sig}^2$.
It is therefore sufficient to derive these in both cases and then compare. \\

\begin{figure}[t]
\centering
\includegraphics[width=\columnwidth]{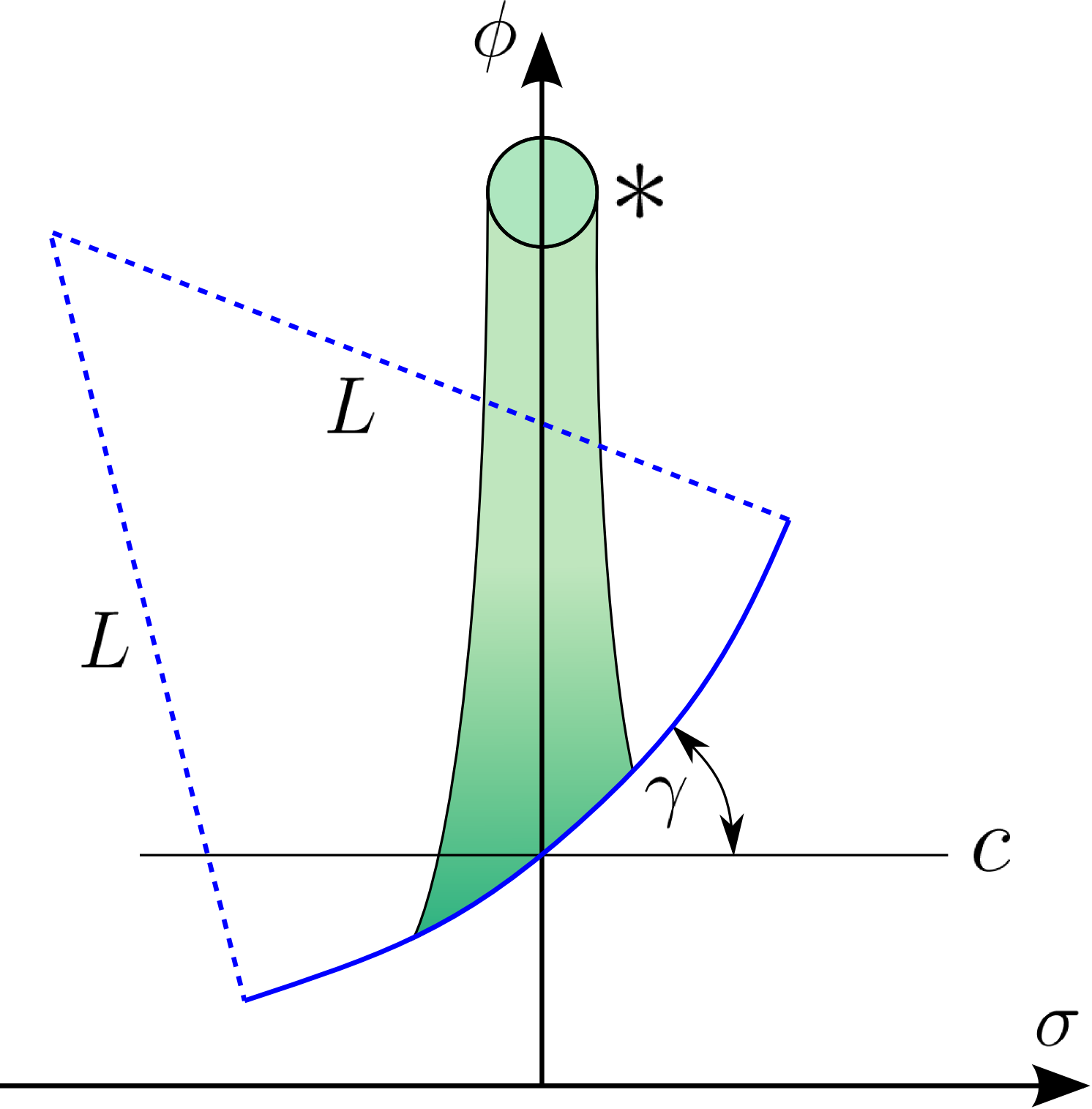}
\caption{The geometry of the {\sc iei} scenario. The bundle (shaded green) begins at horizon crossing `$*$' with perturbations in both fields
(shown schematically as a circle). The bundle then evolves along the $\phi$ direction, but its width is able to vary.
Inflation ends when the bundle hits the curved hypersurface (solid blue line) which has a Gaussian curvature $K = 1/L$.
This hypersurface intercepts a nearby uniform density hypersurface `$c$' at an angle $\gamma$.}
\label{fig:iei}
\end{figure}

\noindent \textbf{\textit{Inhomogeneous end of inflation.}} 
As discussed above, $\zeta_{\rm reh}$ is generated in the {\sc iei} case by the bundle evolving from a hypersurface of uniform density
to a hypersurface with an arbitrary geometry. To compute the bispectrum, we need to specify this geometry to second order.
The first order description is simply the angle between the hypersurfaces, which we denote $\gamma$. 
The second order piece is the rate of change of this angle across the bundle width $\d \gamma / \d \sig^c$ which is non-zero if the end of inflation
hypersurface is curved. It proves more notationally economical to work with the Gaussian curvature $K=\d \gamma / \d \sig^c$ 
which is related to the radius of the arc that defines the end of inflation hypersurface $L$ as $K = 1/L$.
We illustrate this scenario in Fig.~\ref{fig:iei}.
We then calculate $R$ and $N_{,\sig \sig} / N_{,\sig}^2$ as
\bs
\begin{align}
\label{eq:R_iei}
R &= \frac{\ep^*}{\ep^c} \tan^2 \gamma 
\bigg(\frac{\partial \sig^c}{\partial \sig^*}\bigg)^2 \,, \\
\label{eq:fnl_iei}
\!\!\!\!\!\!\!\! \frac{N_ {,\sig \sig}}{N_{,\sig}^2} &\supseteq
-\frac{\sqrt{2 \ep^c}}{\Mpl^{-1}} \bigg[
K \, \cosec^2 \gamma 
+ \cot \gamma 
\frac{\partial^2 \sig^c}{\partial \sig_*^2}
\bigg(\frac{\partial \sig^c}{\partial \sig^*} \bigg)^{\!\!\!-2} \bigg], \!\!\!\!\!
\end{align}
\es
where in the last expression we have omitted terms that cannot produce a large value of $\fnl$. 
For clarity, we shall refer to the two terms in Eq.~\eqref{eq:fnl_iei} as the {\it curvature term} and the {\it non-linear term} with respect to their order of appearance. 
It may appear that there is a divergence in $\fnl$ as $\gamma \to 0$ due to the factors of $\cosec^2 \gamma$ and $\cot \gamma$.
However, these are regularised by the prefactor of $R^2 \propto \tan^4 \gamma$ in the expression for $\fnl$. \\

\noindent \textbf{\textit{Modulated reheating.}} 
In this case one calculates $\zeta_{\rm reh}$ by integrating the fluid equations of 
motion $\d \rho / \d N = -3(1+\omega) \rho$, where 
$\omega=0$ before the modulated reheating hypersurface and $\omega = 1/3$ afterwards.
These have exact solutions, which one can Taylor expand as
\be
\label{eq:mr_zeta_sig}
\zeta_{\rm reh} = -\frac{1}{6} \ln \bigg(1 + \frac{\Gamma_{,\sig}}{\Gamma} \bigg|_c \delta \sig_c + \frac{1}{2} \frac{\Gamma_{,\sig \sig}}{\Gamma} \bigg|_c \delta \sig_c^2 + \dots \bigg) \,.
\ee
We note that in this equation $\Gamma_{,\sig} = \d \Gamma / \d \sig$ is a {\it local} derivative, defined 
on the hypersurface `$c$', rather than the {\it bilocal} quantity
$\partial \Gamma / \partial \sig^*$ that is otherwise commonly employed in the literature. 
We now expand the logarithm to yield
\bs
\begin{align}
N_{,\sig} &= -\frac{1}{6} \frac{\Gamma_{,\sig}}{\Gamma}\bigg|_c \frac{\partial \sig^c}{\partial \sig^*}, \\
\!\!\!\!\!\!\!\!\! N_{,\sig \sig} &= -6 N_{,\sig}^2 \bigg(
\frac{\Gamma_{,\sig \sig} \Gamma}{\Gamma_{,\sig}^2}\bigg|_c \!\!\!- 1\bigg) 
+N_{,\sig} \frac{\partial^2 \sig^c}{\partial \sig_*^2} \bigg(\frac{\partial \sig^c}{\partial \sig^*}\bigg)^{\!\!-1}\!\!\!.\!
\end{align}
\es
These results apply if the argument of the logarithm in 
eq.~\eqref{eq:mr_zeta_sig} is near to unity, which requires 
$N_{,\sig} \ll 10^{5}$ and $N_{,\sig \sig} \ll 10^{10}$ for consistency.
We note that these constraints must be true if $\zeta$ is to be treated perturbatively.

It is not immediately obvious how to express the {\sc mr} scenario in 
a geometric way. To achieve a homogeneous description that does not
depend on the absolute value of $\Gamma$, we choose to define 
the angle $\gamma$ using the ratio of $\Gamma(\sig)$ to $\Gamma|_c$
where $\Gamma|_c$ is a reference constant.
To maximise the similarity with the {\sc iei} case we include a number of other $\O(1)$ coefficients in the definition of $\gamma$,
which we define as the angle between the function $\frac{1}{6} \Mpl \sqrt{2 \ep^c} \Gamma|_c^{-1} \Gamma(\sig)$
and the $\sig$ axis. Note that $\ep^c = 3/2$. This definition yields
$\Gamma_{,\sig} |_c = 6 \Gamma |_c \Mpl^{-1} ( 2\ep^c)^{-1/2} \tan \gamma$
and
$\Gamma_{,\sig \sig} |_c = 6 \Gamma |_c K \Mpl^{-1} (2 \ep^c)^{-1/2} \sec^2 \gamma$
where we have again used the Gaussian curvature $K = \d \gamma / \d \sig^c$.
With these definitions we then find $R$ and $N_{\sig \sig} / N_{,\sig}^2$ as
\bs
\begin{align}
\label{eq:R_mod}
R &= \frac{\ep^*}{\ep^c} \tan^2 \gamma \bigg(\frac{\partial \sig^c}{\partial \sig^*}\bigg)^2 \,, 
\\
\label{eq:fnl_mod}
\!\!\!\!\!\!\!\! \frac{N_{\sig \sig}}{N_{,\sig}^2} &\supseteq
6 - \frac{\sqrt{2 \ep^c}}{\Mpl^{-1}} \bigg[ K \cosec^2 \gamma 
+ \cot \gamma \frac{\partial^2 \sig^c}{\partial \sig_*^2} \bigg(\frac{\partial \sig^c}{\partial \sig^*}\bigg)^{\!\!-2} \bigg]. \!\!\!
\end{align}
\es
We now see the motivation for our definition of $\gamma$: it makes
$R$ and all but one of the terms in $N_{\sig \sig} / N_{,\sig}^2$ functionally identical to that found for the {\sc iei} case. 

Physically, the difference between the {\sc mr} and {\sc iei} scenarios lies in the behaviour of the inflaton
at the time of its reheating. In the {\sc mr} case, $\phi$ is oscillating and so inflation has already ended, whereas in the {\sc iei}
case, $\phi$ is undergoing slow-roll evolution and inflation is still in progress.
By writing the formulae for these scenarios in the above forms we demonstrate that these scenarios are
very similar, involving identical physical processes, although their predictions are not identical.\\

\noindent \textbf{\textit{How to generate $\bm{|\fnl|\geq \O(1)}$.}} 
From the above expressions we see that there are three ways to generate $|\fnl| \geq \O(1)$:
\begin{enumerate}
\item A hypersurface with non-zero curvature, $K \neq 0$. 
\item Non-linear evolution of isocurvature, $\partial^2 \sig^c / \partial \sig_*^2 \propto {\xi_{\sig \sig \sig}^*}^2 \neq 0$.
\item Specific to the {\sc mr} scenario, one finds $\fnl \to 5$ in the limit of large $R$ (this result may be modified if $K$ takes large values).
\end{enumerate}

A very important point to note is that $R$ depends on three quantities: the angle $\gamma$, the evolution of the bundle width $\partial \sig^c / \partial \sig^*$
and the ratio of initial to final field velocities $\ep^* / \ep^c$.
If we were to consider the {\sc iei} scenario and ignore two of these degrees of freedom such that
$\delta \sig^c = \delta \sig^*$ and $\ep^c = \ep^*$, then the only way to obtain a large $\fnl$ is for $K$ or ${\xi_{\sig \sig \sig}^*}^2$ to be much larger than $(2 \ep^c)^{-1/2}$.
This either requires a large curvature of the reheating hypersurface or a total violation of slow-roll.
However, if we allow for $\delta \sig^c \gg \delta \sig^*$ or $\ep^c \ll \ep^*$ then it is quite possible for $R$ to take a significant value even in the limit where $\gamma$ is small.
One interesting effect of this is that it allows us to generate large values of $\fnl$ over a much wider region of parameter space,
without requiring such large values of $K$ or ${\xi_{\sig \sig \sig}^*}^2$.
This leads us to make a preliminary inference: Although a significant region of the parameter 
space for these models will produce a small 
bispectrum signal, a non-negligible region of parameter space will be ruled out 
by Planck's bounds on the bispectrum. We shall show this concretely by considering
two representative examples in \S\ref{sec:models}.

A second important point to note is that the value of $\ep^c$ differs between the two scenarios, 
being a slow-roll parameter for the {\sc iei} scenario and $\ep^c = 3/2$ in the {\sc mr} case.
This has two competing effects for the generation of large $\fnl$: On the one hand, $N_{\sig \sig} / N_{,\sig}^2$ is proportional to $\sqrt{2 \ep^c}$ 
and so the larger value of $\ep^c$ in the {\sc mr} scenario makes it easier to generate large $\fnl$. On the other hand, the ratio $\ep^* / \ep^c$ appearing
in $R$ is now suppressed in the {\sc mr} scenario and so this will act to mitigate the opportunity for large $\fnl$. 
There does not appear to be a clear dominance of either of these effects, so we shall 
simply accept that this leads to different effects in the predictions of the two scenarios.

\section{Representative examples}
\label{sec:models}

The previous section discussed the physical mechanisms that can lead to a large $\fnl$ in the {\sc iei} and {\sc mr} scenarios.
We now wish to concretize these statements by applying observational constraints. For this to be possible we need to 
make some choices about the form of the inflationary potential that defines the dynamics of the inflaton $\phi$. \\

\noindent \textbf{\textit{Inhomogeneous end of hybrid inflation.}} 
For the {\sc iei} scenario we shall pick a typical hybrid scenario with a small initial velocity $\ep^* = 10^{-4}$ at horizon crossing
and choose the parameters $\eta_{\phi \phi}$, $\eta_{\sig \sig}$ and $\xi_{\sig \sig \sig}^2$ to be
constants.
We can then find $\ep^c$ by integrating the equation $\d\ep / \d N = 4 \ep^2 - 2 \ep \eta_{\phi \phi}$ over 60 efolds to obtain
\be
\label{eq:ep_iei}
\ep^c (\eta_{\phi \phi}) = \ep^* \bigg[ \frac{2 \ep^*}{\eta_{\phi \phi}} \bigg( 1 - e^{120 \eta_{\phi \phi}} \bigg) + e^{120 \eta_{\phi \phi}} \bigg]^{-1} .
\ee
Requiring $\ep^c <0.1$ to maintain slow-roll consistency we obtain the bound $\eta_{\phi \phi} > -0.043$.
We then impose further slow-roll bounds as $\eta_{\phi \phi} < 0.1$ and $-0.1 < \eta_{\sig \sig} < 0.1$.

The next constraint we can apply is that of the spectral index. Since the constraint provided by
Planck is very tight, we assume for simplicity that our model exactly reproduces 
the best-fit Planck value, which allows us to write the function $R$ in terms of $\eta_{\phi \phi}$ and $\eta_{\sig \sig}$ as
\be
\label{eq:R}
R(\eta_{\phi \phi},\eta_{\sig \sig}) = \frac{2 \eta_{\phi \phi} - 6 \ep^* - (n_\zeta - 1)}{(n_\zeta - 1) + 2 \ep^* - 2 \eta_{\sig \sig}} \,.
\ee
Positivity of $R$ requires that the numerator and denominator of Eq.~\eqref{eq:R} are either both
positive or negative.
This splits the $\{\eta_{\sig \sig}, \eta_{\phi \phi}\}$ phase space into four regions, only two of which are viable. 
There is also a constraint arising from the upper bound on the tensor-scalar ratio, although the small value of $\ep^*$ means that this is satisfied
for the whole of the $\{\eta_{\sig \sig}, \eta_{\phi \phi}\}$ space.

\begin{figure}[b]
\centering
\includegraphics[width=\columnwidth]{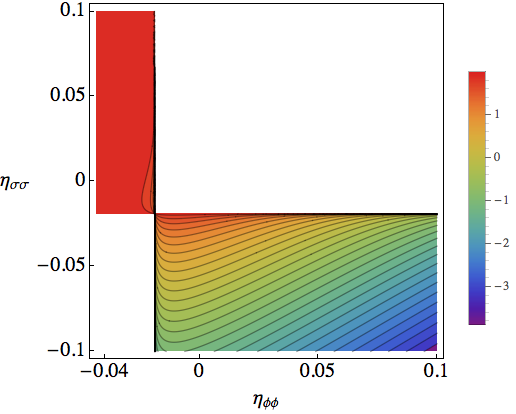}
\caption{Values of $\gamma$ in degrees over the phase space for hybrid inflation. 
The plot is logarithmic, so $\gamma = 10^x$ where $x$ may be read from the legend. 
The two white regions are excluded by the requirement of a viable spectral index.
Each phase space point has a precise value of $\gamma$ in order to obtain the correct value for the spectral index.}
\label{fig:g_iei}
\end{figure}
We may then calculate $\gamma(\eta_{\phi \phi},\eta_{\sig \sig})$ by inverting 
Eq.~\eqref{eq:R_iei} and substituting for $\ep^c$ using Eqs.~\eqref{eq:ep_iei}. The evolution of isocurvature is given by
Eqs.~\eqref{eq:isocurvature1}-\eqref{eq:nonlinear1}. For this hybrid model, these formulae simplify because 
$V_{,\sig \sig}$ and $V_{,\sig \sig \sig}$ are constant by virtue of $\sig$ being fixed at zero.
Furthermore, the dominant vacuum term ensures that $H$ is roughly constant during inflation and so 
Eqs.~\eqref{eq:isocurvature1}-\eqref{eq:nonlinear1} simplify as
\bs
\begin{align}
\label{eq:iei_iso1}
\frac{\partial \sig^c}{\partial \sig^*} &\simeq e^{-60 \eta_{\sig \sig}^*} \,, \\
\label{eq:iei_iso2}
\frac{\partial^2 \sig^c}{\partial \sig_*^2} &\simeq - \frac{\partial \sig^c}{\partial \sig^*} \bigg(1-\frac{\partial \sig^c}{\partial \sig_*} \bigg) \frac{{\xi_{\sig \sig \sig}^*}^2}{\Mpl \sqrt{2 \ep^*} \eta_{\sig \sig}^*}
\,,
\end{align}
\es
where we have presumed 60 efolds of observable inflation.
Combining these ingredients we find $\gamma(\eta_{\phi \phi},\eta_{\sig \sig})$ which we then plot in Fig.~\ref{fig:g_iei}. 
The region with more positive $\eta_{\sig \sig}$ requires predominantly large angles $\gamma$, whereas the region with more
negative $\eta_{\sig \sig}$ is valid for a wide range of smaller angles.

\begin{figure}[t]
\centering
\includegraphics[width=\columnwidth]{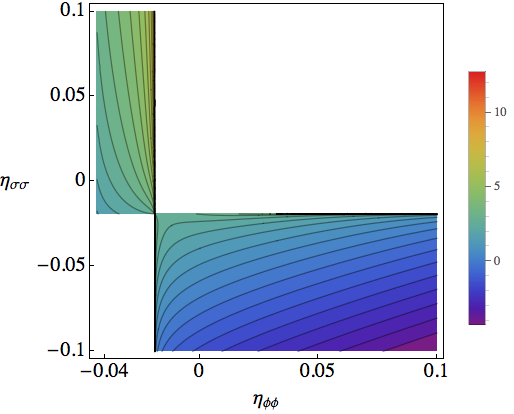}
\includegraphics[width=\columnwidth]{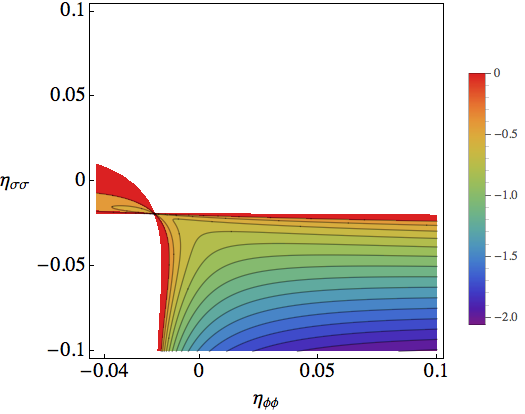}
\caption{{\it Top pane}: Upper bounds on $|K \, \Mpl|$ for $K<0$ on a logarithmic scale ($|K\, \Mpl|=10^x$ where $x$ relates to
the legend shown). Again, the two white regions are excluded by the requirement of a viable spectral index. \newline
{\it Lower pane}: Upper bounds on $\xi_{\sig \sig \sig}^2$ for $\xi_{\sig \sig \sig}^2 >0$, on a similar logarithmic axis.}
\label{fig:K_iei}
\end{figure}
We can now consider $\fnl$ by looking at Eq.~\eqref{eq:fnl_iei}. It is natural to split this calculation into two parts,
firstly that with $K \neq 0$ and $\xi_{\sig \sig \sig}^2=0$, and secondly with $\xi_{\sig \sig \sig}^2 \neq 0$ and $K=0$.
For the case with non-zero curvature, and working to the Planck 1-sigma bounds on $\fnl$, we obtain slightly different 
but broadly similar constraints on positive and negative curvature $K$. 
We plot the constraint on negative $K$ in Fig.~\ref{fig:K_iei}. This shows a wide range of upper bounds on $|K|$. By comparison with
Fig.~\ref{fig:g_iei}, we see that the regions with the tightest upper bounds on $|K|$ are those with the smallest angles $\gamma$. 
Specifically, when $\eta_{\phi \phi}=0.1$ and $\eta_{\sig \sig}=-0.1$ we find $-4.6 \times 10^{-5} < K \, \Mpl < 1.7 \times 10^{-5}$. 
However, we note that this requires a very small angle of $\gamma = 4.3 \times 10^{-4}$  degrees
and, depending on the model building scenario, this in itself may be fine tuned. 
Alternatively, if we set $\gamma=1$ degree, then the smallest bounds that we can place on $K$ are 
$-2.03 < K \, \Mpl< 0.74$. Both sets of bounds are surprisingly small, demonstrating that the bispectrum 
is able to place significant constraints on the curvature of the reheating hypersurface.

The non-linear term in Eq.~\eqref{eq:fnl_mod} is parametrized by $\xi_{\sig \sig \sig}^2$. 
We may perform an analogous analysis to that above for $K$ and obtain bounds on $\xi_{\sig \sig \sig}^2$,
which again will differ depending on whether this third slow-roll parameter is positive or negative.
The tightest constraints arise for the bottom right-hand corner of the phase space in Fig.~\ref{fig:K_iei}
where we find $0.0032 < \xi_{\sig \sig \sig}^2 < 0.0087$. 
For a larger angle of $\gamma = 1$ degree the tightest constraints that we find are
$-0.034 < \xi_{\sig \sig \sig}^2 < 0.095$.
These latter constraints are weaker than those expected purely from slow-roll. We conclude that the non-linear isocurvature evolution of these models produces a non-Gaussianity that is
consistent with the Planck bispectrum across the vast majority of the parameter range.\\

\noindent \textbf{\textit{Modulated reheating of vanilla inflation.}} 
For modulated reheating, the requirement of a quadratic minimum and a graceful end to inflation are both most easily satisfied by
working with the `vanilla inflation' potential $V(\phi) = \frac{1}{2} m_\phi^2 \phi^2$. We prescribe $60$ efolds of slow-roll inflation to arise between
horizon exit and the time when $\ep^c=3/2$, where we presume that the oscillations begin and that there is no further evolution of isocurvature.
Since we know the form of the potential, $\ep$ and $\eta_{\phi \phi}$ follow straightforwardly following the assumption of slow roll.
Again, the lack of direct coupling between the $\phi$ and $\sig$ fields means that we still expect
$V_{,\sig \sig}$ and $V_{,\sig \sig \sig}$ to remain constant, but the fact that inflation ends gracefully in this model prevents us from
approximating the Hubble parameter to be a constant.
This enters our calculation through the evolution of isocurvature, which we may calculate from Eqs.~\eqref{eq:isocurvature1}-\eqref{eq:nonlinear1} 
under the approximation of slow-roll evolution. We find
\bs
\begin{align}
\frac{\partial \sig^c}{\partial \sig^*} &\simeq e^{-\alpha \eta_{\sig \sig}^*} \,, \\
\frac{\partial^2 \sig^c}{\partial \sig_*^2} &\simeq - \frac{\partial \sig^c}{\partial \sig^*} \frac{{\xi_{\sig \sig \sig}^*}^2}{\Mpl \sqrt{2 \ep^*} \eta_{\sig \sig}^*}
\,,
\end{align}
\es
where $\alpha = -\frac{1}{2} \phi_*^2 \ln (\phi^c / \phi^*)$.
We find $\phi^* = 15.53\, \Mpl$ in order to get 60 efolds of inflation, and by presuming $\ep^c = 3/2$ we find $\phi^c = 2/\sqrt{3}$.
This then gives $\alpha  = 313.6$. This contrasts to the value of $60$ found for the hybrid {\sc iei} case, meaning that we expect more significant evolution of isocurvature in this {\sc mr} model.
For the {\sc mr} scenario one would also expect some isocurvature evolution to occur during 
the dust-like phase where the inflaton is oscillating, but we ignore this contribution by assuming that the inflationary efoldings dominate.

\begin{figure}[b]
\centering
\includegraphics[width=\columnwidth]{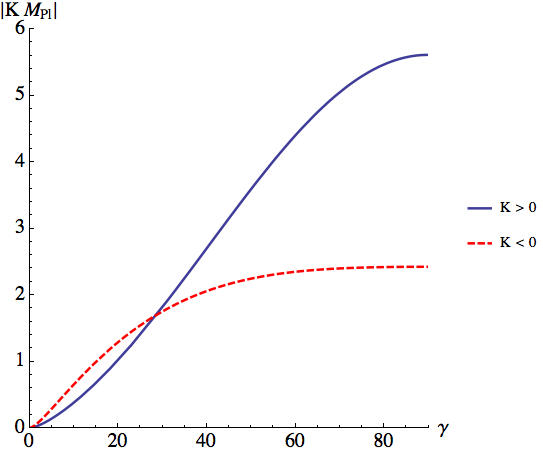}
\includegraphics[width=\columnwidth]{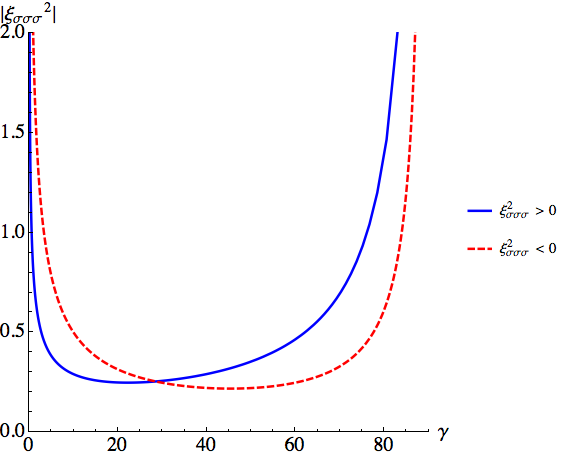}
\caption{{\it Top pane}: Upper bounds on $|K\, \Mpl|$ plotted against $\gamma$, showing reasonable constraints across all angles, but particularly tight constraints at small angles.\newline
{\it Lower pane}: Upper bounds on $|\xi_{\sig \sig \sig}^2|$. 
}
\label{fig:mr}
\end{figure}
We also anticipate a minor tension with the tensor-scalar ratio for this class of models, but since this is only marginal and furthermore since our
aim here is to show a simple illustrative example for constraints on $\fnl$, we shall ignore this issue.

Since $\eta_{\phi \phi}^*$ is prescribed by the model, $\eta_{\sig \sig}^*$ is now the only arbitrary degree of freedom.
One finds $R$ and $\gamma$ identically to the method used in the previous case, except in this case they only vary with $\eta_{\sig \sig}^*$.
The requirement of a viable spectral index places an upper bound on $\eta_{\sig \sig}^*$, and we obtain a lower bound
on $\eta_{\sig \sig}^*$ by demanding that $\delta \sig^c < 0.1 \, \Mpl$. This leaves a viable parameter window as
$-0.031 < \eta_{\sig \sig}^* < -0.011$.

Bounds on $K$ and $\xi_{\sig \sig \sig}^2$ then follow by taking each term separately and these are shown in Fig.~\ref{fig:mr}.
We see that $K$ is bounded as $-2.42 < K \, \Mpl< 5.61$ {\it over the full range of angles $\gamma$}, with much tighter bounds at low angles.
Specifically, for $\eta_{\sig \sig} = -0.031$ we find $-3.36 \times 10^{-5}< K\, \Mpl  < 1.26 \times 10^{-5}$ which occurs when $\gamma = 0.016$ degrees.
Again we find that the non-linear term yields constraints that are weaker than those required for slow-roll consistency.
The tightest bounds are $-0.22 < \xi_{\sig \sig \sig}^2< 0.25$ which requires $22< \gamma <45$ degrees.

\section{Conclusions} \label{sec:conclusions}

A key question concerning inflation is how to narrow down the plethora of models that have been
proposed, and in particular how this is affected by the bounds on non-Gaussianity recently
obtained by Planck. Given the compatibility of Planck data with small non-Gaussianity, it could be tempting to interpret this
as an indication that the models underpinning inflation are likely to be the simplest  models compatible with the data,
namely canonical single field models, as indeed has been recently argued by a number of authors.

To assess how reasonable this interpretation is, we have considered two-field models as the simplest 
multi-field generalisations, and have looked at ways in which these models can behave differently from their single field counterparts. 
Given that the two-field slow-roll inflation and the curvaton scenario
have been studied recently, we have concentrated on the Modulated Reheating ({\sc mr})
and the Inhomogeneous End of Inflation ({\sc iei}) scenarios. Employing a geometrical approach, 
we have shown that these two scenarios are very similar, though not identical.

Generalising previous work to allow for arbitrary evolution of the isocurvature perturbations,
we have shown that there are two physical mechanisms that can produce $\fnl$ in 
excess of the Planck bispectrum bounds: Firstly, reheating may occur on a curved hypersurface, 
leading to a contribution to $\fnl$ that is proportional to the Gaussian curvature $K$ of 
the reheating hypersurface. The capacity for this term to generate $|\fnl| \gg 1$
is significantly enhanced for models for which the isocurvature perturbations grow.
Secondly, one obtains a contribution to $\fnl$ from the non-linear growth of the isocurvature perturbations,
which is proportional to the third slow-roll parameter $\xi_{\sig \sig \sig}^2$.
We have shown how both these effects can be parametrized in terms of the angle $\gamma$
that measures the relative orientation of the reheating hypersurface from one of uniform density.

To concretize the degree of fine tuning required by these scenarios to produce a large bispectrum, we have 
considered two representative examples, hybrid inflation with an 
inhomogeneous ending, and vanilla inflation with modulated reheating.
By scanning the relevant parameter spaces, we have placed bounds on $K$ or $\xi_{\sig \sig \sig}^2$,
by treating each source of the bispectrum independently.
These bounds are fully consistent with the bounds on the spectral index.

For these particular models we find that Planck bounds can eliminate significant ranges of possible values of $K$, 
with this effect being most striking in the {\sc mr} model where we find $-2.42 < K \, \Mpl< 5.61$ as the {\it weakest} bound in the parameter space under consideration.
These bounds become much tighter for values of $\gamma$ nearer to zero, such as the bound
$-3.36 \times 10^{-5}< K \, \Mpl < 1.26 \times 10^{-5}$ which occurs when $\eta_{\sig \sig} = -0.031$ and requires $\gamma = 0.016$ degrees.
For the {\sc iei} model we obtain a wide range of constraints, with $|K\, \Mpl|<\O(1)$ for a significant region of the parameter space,
of in more restricted scenarios we obtain tighter bounds such as 
$-4.6 \times 10^{-5} < K \, \Mpl< 1.7 \times 10^{-5}$ which applies for the case with $\eta_{\phi \phi}=0.1$ and $\eta_{\sig \sig}=-0.1$,
although we note that this case requires $\gamma = 4.3 \times 10^{-4}$ degrees.
These examples show clearly that Planck spectral index and bispectrum bounds are definitely not able to rule out
these multi-field scenarios, although we do find that they are able to place some significant constraints
on the geometry of the reheating hypersurface.

We do not find the same behaviour for the contribution to $\fnl$ that is mediated by non-linear growth of the isocurvature perturbations
because $\xi_{\sig \sig \sig}^2$ is constrained to have small values by slow-roll. We find that this `non-linear' contribution is totally unconstrained 
for vanilla inflation with additional {\sc mr} effects, and for the hybrid {\sc iei} model it can only be constrained
for very small regions of the parameter space which coincide with very small values of the angle $\gamma \sim 10^{-4}$ degrees.
This is similar to the scenario which we encountered for models of two-field slow-roll inflation or in curvaton-type models,
where only very restricted regions of initial condition and parameter space lead to a large non-Gaussianity.
As a result, such models remain compatible with observational data for the vast majority of their parameter spaces.

In summary, we find that the simplest representative two-field inflationary models are easily consistent with
the tighter bounds that Planck has placed on the spectral index and the bispectrum,
since they only generate a large bispectrum for certain regions of their parameter space.
These results, combined with the motivation for considering multi-field models coming from candidate theories of 
fundamental interactions, suggest that inflation sourced by more than one scalar 
field remains an important possibility.\\

\noindent {\bf Acknowledgements}.
We would like to thank Chris Byrnes, David Seery and David Wands for their very helpful and constructive comments.
JE was supported by the Science and Technology Facilities Council grant ST/I000976/1 and DJM and RT by the 
Science and Technology Facilities Council grant ST/J001546/1.

\bibliography{postplanck}

\end{document}